\documentclass{article}

\usepackage{PRIMEarxiv}

\usepackage[utf8]{inputenc} 
\usepackage[T1]{fontenc}    
\usepackage{hyperref}       
\usepackage{url}            
\usepackage{booktabs}       
\usepackage{amsfonts}       
\usepackage{nicefrac}       
\usepackage{microtype}      
\usepackage{lipsum}
\usepackage{fancyhdr}       
\usepackage{graphicx}       
\graphicspath{{media/}}     
\usepackage{times}

\usepackage{soul}
\usepackage{url}
\usepackage{graphicx}
\usepackage{booktabs}
\usepackage{amsmath}
\usepackage{bm}
\usepackage{color}
\usepackage{pgfplots}
\usepackage{subcaption}
\usepackage{wrapfig, blindtext}
\usepackage{comment}

\usepackage{algorithmic}
\usepackage[ruled,vlined,linesnumbered]{algorithm2e}

\usepackage{balance}

\definecolor{darkgreen}{rgb}{0, 0., 0}
\definecolor{green1}{rgb}{0, 0., 0}
\definecolor{new}{rgb}{0, 0.0, 0}
\title{Intelligent Autonomous Intersection Management}

\author{
Udesh Gunarathna\\
University of Melbourne\\
Melbourne\\
Australia\\
pgunarathna@student.unimelb.edu.au\\
\And
Shanika Karunasekara\\
University of Melbourne\\
Melbourne\\
Australia\\
karus@unimelb.edu.au\\
\And
Renata Borovica-Gajic\\
University of Melbourne\\
Melbourne\\
Australia\\
renata.borovica@unimelb.edu.au\\
\And
Egemen Tanin\\
University of Melbourne\\
Melbourne\\
Australia\\
etanin@unimelb.edu.au\\
}

\begin{document}
\maketitle

\begin{abstract}
Connected Autonomous Vehicles will make autonomous intersection management a reality replacing traditional traffic signal control. Autonomous intersection management requires time and speed adjustment of vehicles arriving at an intersection for collision-free passing through the intersection. Due to its computational complexity, this problem has been studied only when vehicle arrival times towards the vicinity of the intersection are known beforehand, which limits the applicability of these solutions for real-time deployment. \textcolor{darkgreen}{To solve the real-time autonomous traffic intersection management problem, we propose a reinforcement learning (RL) based multiagent architecture and a novel RL algorithm coined multi-discount Q-learning. In multi-discount Q-learning, we introduce a simple yet effective way to solve a Markov Decision Process by preserving both short-term and long-term goals, which is crucial for collision-free speed control. Our empirical results show that our RL-based multiagent solution can achieve near-optimal performance efficiently when minimizing the travel time through an intersection.}
\end{abstract}


\section{Introduction}

AIM is a paradigm proposed by Dresner \emph{et al.}~\cite{AIM_definition} to replace the traditional traffic signal control. In AIM, each CAV arriving towards an intersection reserves a time to reach the intersection crossing point via an intersection manager. \textcolor{darkgreen}{Then, each CAV's speed is controlled to adhere to the schedule while guaranteeing safety}\footnote{\textcolor{darkgreen}{In traditional AIM, vehicles traverse through the intersection on a First Come, First Served basis. In this research, we optimize the traversing order to reduce the congestion further, e.g. support platooning.}}.
Due to the ability of AIM to reduce the congestion at intersections~\cite{AIM_survey}, AIM has been widely studied~\cite{AIM_1,AIM_5}. However, most problem formulations assume that the arrival times of vehicles to the vicinity of intersection are known \emph{a priori}. This assumption does not hold when dealing with real-time traffic, which limits the applicability of these solutions to real life scenarios~\cite{polling-system}. 

For AIM to be applicable to real life traffic control, the speed of the vehicles has to be computed in real-time, and the computational time for this plays a critical role in the feasibility of AIM.
Previous efforts exhibit high computational time~\cite{np-hard} as they rely on mathematical programming or analytical methods.
The impact of high computational time is two fold. First, suppose the intersection controller takes a long time to compute scheduled times. In that case, the positional difference of CAVs before and after the computations, poses a significant safety risk of the vehicle crashing due to the given schedule times not being feasible anymore. Second, if the time for computing CAV speed for the trajectory is high, the remaining time after the computation may not be sufficient to reach the intersection crossing point precisely at the scheduled time. 

Recent work \cite{polling-system} develops a stochastic solution to the problem, assuming vehicle arrival times are not known beforehand. However, a part of their solution employs linear programming (LP) for every CAV's arrival computation, which is prohibitively expensive, as further demonstrated in Section~\ref{sec-exp-results}. Further, their method is only applicable to intersections with single lane roads without turning directions.

\textcolor{darkgreen}{Our work fills the missing research gap by providing a computationally efficient, deployable, multiagent solution based on Reinforcement Learning (RL) for AIM. 
Our solution consists of two main sets of agents. The first type of agent is a polling-based coordinating agent (intersection controller) positioned at the intersection. The second component is a set of distributed RL agents, which are assigned on a per vehicle basis. The coordinating agent communicates with the RL agents to schedule time intervals to reach the intersection for each CAV that is within a certain distance from the intersection. The coordinating agent uses a novel polling algorithm to handle multi-lane intersections with multiple turning directions, overcoming thereby a major limitation of previous work~\cite{polling-system}. Once the coordinating agent provides a time schedule, an RL agent controls each vehicle's speed to adhere to the coordinating agent's time schedule.} The advantage of such an approach is that once an RL agent is trained offline, decision-making can be done much faster online. This avoids the computation overhead incurred by LP-based techniques.  

The learning task for the RL agent is two fold: (1) learn to control a vehicle's trajectory to reach the intersection precisely at the scheduled time, and (2) keep a safe distance from the vehicle in front. Keeping a safe distance is a task with a short-term goal, because the front vehicle can change its driving behaviour (the driving behavior of an RL agent or a human) in short time-intervals. In contrast, reaching the intersection at a scheduled time requires long-term planning because successfully reaching the intersection is only determined at the end of the trajectory. Combining such two learning problems into a single task, as shown in previous work \cite{Mo-MDP} will only learn one of the problems, because each learning problem contains an objective with a different time-horizon.
Traditional RL algorithms such as Q-learning use a fixed parameter named \emph{discount factor} to control the length of the time-horizon. Using a fixed discount factor focuses on learning either the short-term or long-term task successfully~\cite{discount-factor-1}, and fails when both short-term and long-term tasks need to be learned simultaneously, hence being unsuitable for our task. We propose a novel RL algorithm coined \emph{multi-discount Q-learning} to achieve short-term goals, while following a long-term goal in a single Markov Decision Process. We believe our proposed method is applicable to other problem-domains that exhibit a mix of long-term and short-term goals, such as robotics~\cite{robotics}. 

Our contributions are four-fold: \textcolor{darkgreen}{(1) We propose a computationally efficient multiagent solution for AIM. (2) We introduce a novel reinforcement learning algorithm that can effectively learn multiple learning tasks. (3) We propose a novel polling algorithm to handle multi-lane intersections with multiple turning directions.} (4) We demonstrate the effectiveness and efficiency of our approach against several baselines using microscopic traffic simulations.  


\section{Related Work}
\subsection{Autonomous Intersection Management}
There are two inter-dependent sub-problems that have been studied in the literature to optimize the AIM; (1) find a distinct time-schedule for each incoming vehicle to arrive at the intersection, and (2) find a vehicle trajectory such that a vehicle arrives at the intersection exactly at the schedule time\footnote{If the objective is to optimize the throughput then the vehicle trajectory should reach the intersection at the maximum speed as well~\cite{max-speed}.}. The existing work can be divided into two main categories based on the proposed solutions to the aforementioned two problems.

The first category of work optimizes the time-schedule (sub-problem (1)) as a scheduling problem and then computes vehicle trajectories which adhere to the optimized time-schedule \cite{AIM_5,AIM_6,AIM_7}. The scheduling optimization is NP-hard \cite{np-hard}, which makes it computationally expensive. Another drawback of this kind of approach is that all the arrival times of vehicles need to be known \emph{beforehand} to optimize the time-schedule. 
This property limits their applicability to real-time settings where vehicle arrival times are stochastic.  

The second category of work focuses on finding a safe trajectory or fastest trajectory as a solution to sub-problem (2), whilst employing a heuristic to compute the time-schedule to sub-problem (1)~\cite{AIM_1,AIM_4,AIM_analytical}. Finding the optimal trajectory is however prohibitively expensive in real-time when using a method like linear programming, as demonstrated in our experiments. For example, Au \emph{et al.}~\cite{AIM_analytical} uses an analytical method to find the trajectory using a set-point schedule and a bisection method. However, to simplify the search space when deciding on the trajectory their work does not consider the maximum velocity, which leads to a sub-optimal trajectory. Even though these approaches are able to compute trajectories, they do not optimize sub-problem (1). Thus, they do not optimize the throughput nor reduce waiting time at intersections. In contrast, our objective is to maximize the throughput at the intersection.

Recent work \cite{polling-system} proposed a solution to optimize the throughput in a stochastic setting using a polling system and linear programming. The polling system is used to optimize the time-schedule for each vehicle. Then, a linear program is solved for each vehicle to find its trajectory. These linear programs need to be computed sequentially (centralized). This means that the linear program for the first vehicle arriving at the intersection should be computed first, and then the next vehicle, and so on. Although this approach can successfully solve the stochastic AIM problem, computational time required for linear programming hinders the applicability of this solution for real-time usage. On the contrary, we propose a distributed learning-based solution, which enables real-time AIM. 

\vspace{-0.5em}
\subsection{RL with Variable Discounting}

Learning a task is difficult when there are objectives with different time scales. The time scale of a task is directly impacted by the discount factor of RL agents (e.g. Q-learning or SARSA) \cite{discount-factor-1}. Thus, most past efforts focus on changing the discount to learn tasks with multiple time scales. Edwards \emph{et al.} \cite{separate-discount-factors} extends the LP formulations of \cite{discount-factor-1} and propose a multiple discount SARSA algorithm by considering the reward as a vector and using a separate action-value function (Q-function) for every sub-task. Human intervention is then needed to find the best policy among these sub-tasks. Burda \emph{et al.} \cite{intrinsic-discount-factors} follows a similar approach to combine intrinsic and extrinsic rewards. \textcolor{darkgreen}{An automated approach is proposed by Li \emph{et al.}~\cite{aamas_mqdqn} to combine different objectives learned by a set of factored Q-functions using a lexicographic ordering of objectives. Finding such lexicographic ordering of objectives is non-trivial and can be problem-dependent. In the above-mentioned approaches, each sub-task is learned separately. Because of that, the inter-relationship between sub-tasks is ignored, and the number of parameters to be learned is high. In contrast, we propose a simple and memory-efficient method to learn each sub-task using a single action-value function (a single Q-function) whilst preserving the time scales of sub-tasks. As we show in our experiments, our proposed method is able to achieve superior results in achieving both short-term and long-term goals.}

A single Q-function is similarly used with a state-dependent discount function where each state is discounted differently \cite{state-dependent}. Also, a generic hyperbolic discount function is proposed as opposed to the traditional exponential discount function \cite{hyperbolic}. However, these algorithms focus on a single scalar reward, whereas our proposed method extends to multi-objective cases which contain multiple rewards with different temporal objectives, allowing us to preserve the time scale of each objective.

\section{Background} \label{sec-background}
Our proposed architecture uses a polling system to schedule the incoming CAVs, and Q-learning to determine the CAV trajectories. We provide a brief introduction to both.

\textbf{Polling system:} A polling system consists of a single server and a set of queues. Each queue contains a number of customers (in First-In First-Out (FIFO) order). Customers may arrive at the queue in a stochastic order. The server can start serving a customer from any queue. The term \emph{service time} is the time taken to service one customer. Once the server serviced the first customer, the server can select the next customer from any queue. When switching between queues, the server has to wait for an additional time called \emph{switch over time}. The strategy that the server uses to determine from which queue the next customer is selected is called the \emph{policy}. There are several policies in the literature such as \emph{K-limited}, \emph{gated} and \emph{exhaustive}. The definitions of customers, \emph{switch over time} and \emph{service time} related to AIM are described in Section \ref{sec-polling-system}.

\textbf{Q-learning:} In RL, a problem first needs to be formulated as a Markov-decision process (MDP). An MDP consists of a state space $\mathcal{S}$ and an action space $\mathcal{A}$. When an action $a_t \in \mathcal{A}$ is taken  in the current state $s_t \in \mathcal{S}$, at time $t$, the MDP's state changes to $s_{t+1}$ according to the transition probability $\mathcal{T}(s_t,a_t,s_{t+1}) = Pr(s_{t+1}|s_t,a_t)$. The MDP provides a reward  $r_t$ for the transition where $r_t$ is assigned according to $\mathcal{R}(s_t,a_t,s_{t+1})$. An RL agent acting on the MDP consists of a policy $\pi(a|s)$ which describes the agent's behaviour. The policy $\pi(a|s)$ indicates the probability of an agent taking the action $a$ in the state $s$. The objective of the agent is to maximize the \emph{expected reward} $G_t$ starting from any given time step $t$. The expected reward is defined as $G_t = \sum^T_{\tau=t}{\gamma^{\tau-t}r_\tau}$, where $t$ is the current time step, $\gamma$ is the discount factor and $T$ is the time  MDP reaches its terminal state. 

The action-value function (Q-function) for policy $\pi$ stores the expected reward by taking the action $a$ in the state $s$ defined as: $Q^{\pi}(s,a) = \mathop{\mathbb{E}}[G_t|s_t=s, a_t=a,\pi]$. 
Q-learning approximates the optimal Q-function iteratively by observing the transitions ($s_t, a_t, s_{t+1}, r_t$) at every time step when  $\mathcal{T}$ is unknown. Considering the reward from $n$ number of steps we get the following n-step Q-learning equation.
\vspace{-0.5em}
\begin{equation} \label{n-step-q-learning}
    Q(s_t,a_t) \gets Q(s_t,a_t) + \alpha[ \sum_{\tau=0}^{n-1} \gamma^\tau r_{t+\tau} + \gamma^n \max_{a^\prime} Q(s_{t+n}, a^\prime) \\ - Q(s_t,a_t)]
\end{equation}

\textbf{Deep Q-learning:} 
Deep Q-learning (DQN)~\cite{deep-Q} uses a neural network to approximate the tabular Q-function when the number of state-actions is large. The deep neural network is represented as $Q(s,a;\theta)$ where $\theta$ is the set of parameters of the neural network. DQN learns the Q-function minimising the following loss function w.r.t $\theta$. 
\vspace{-0.5em}
\begin{equation} \label{deep-q-learning}
    L_t(\theta_t) = \mathbb{E}_{(s,a,r,s^\prime) \sim \mathcal{D}} [(y_t^{DQN} - Q(s,a;\theta_t))^2]
\end{equation}
\begin{equation} \label{target}
    y_t^{DQN} = r + \gamma \max_{a^\prime} Q(s^\prime, a^\prime; \theta_t^-)
\end{equation}

where  $y_t^{DQN}$ is the \emph{target function} and $\theta^-$ parameters represent a separate neural network that keeps its parameters frozen for a number of time steps while ($\theta$) parameters are optimized. After a certain number of time steps $\theta$ parameters are copied to $\theta^-$. 

\section{Problem Definition} \label{sec-prob-def}
\begin{figure}[t]
\begin{subfigure}{0.485\columnwidth}
    \centering
    \includegraphics[width=0.94\columnwidth]{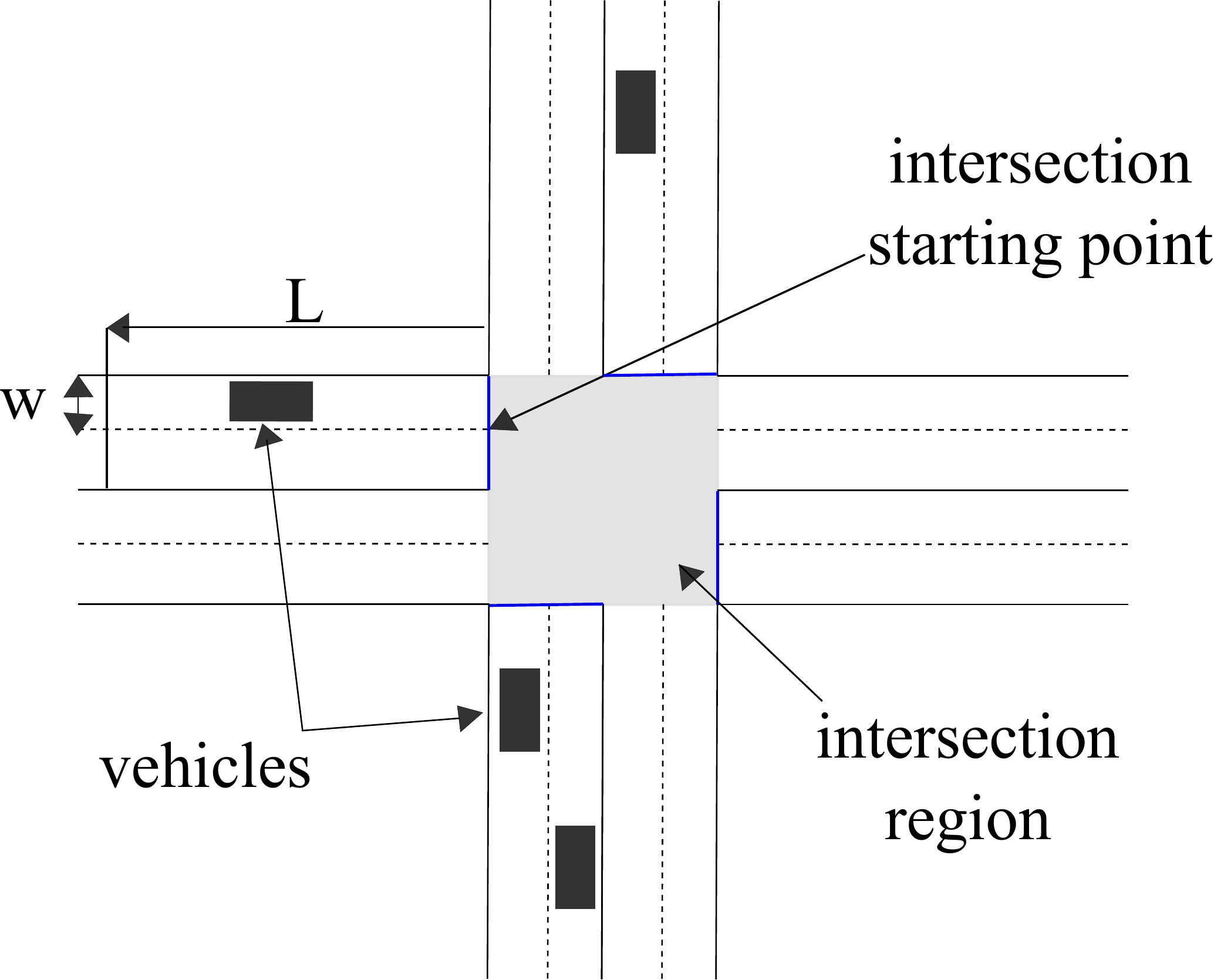}
    \caption{A four legged intersection with vehicles represented with black rectangles. The control region for one road segment is shown with length $L$. The intersection region is shown in gray and the intersection starting point for road segments are shown in blue lines.}
    \label{fig:Intersection}
\end{subfigure}\hspace{1.5mm}\begin{subfigure}{0.485\columnwidth}
    \includegraphics[width=0.77\columnwidth]{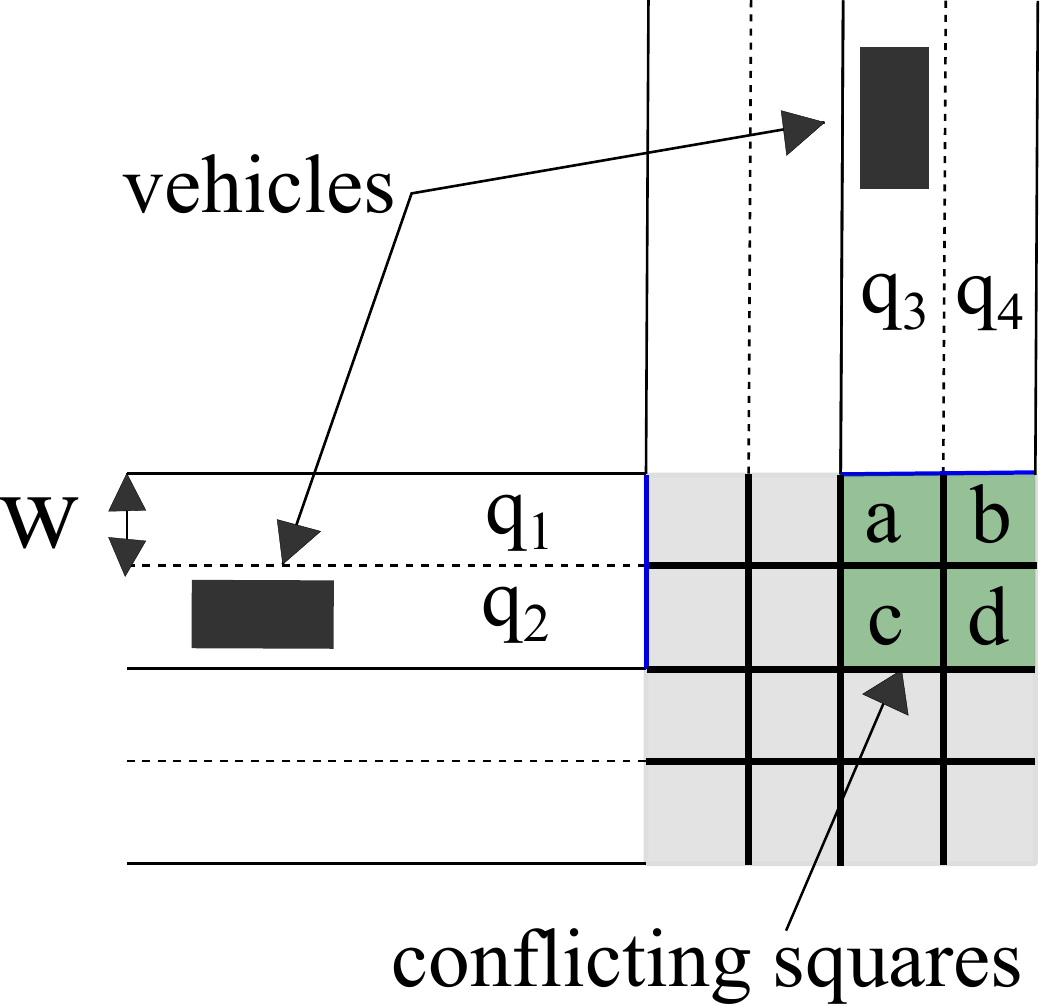}
    \caption{\textcolor{darkgreen}{The figure shows two 2-lane roads segments where each lane is represented as a queue. The green points show four possible conflicting squares where two vehicles from 2 lanes could collide, if vehicles coming from $q_1$ to $q_4$ are going straight through the intersection.}}
    \label{fig:crossing-points}
\end{subfigure}
    \centering
    \vspace{-0.6em}
    \caption{Multi-lane signal-less intersection}
    \label{fig:my_label}
    \vspace{-5mm}
\end{figure}

\textcolor{darkgreen}{Let us assume that an intersection I is connected with K road segments. A vehicle $i$ travelling towards the intersection from road segment $k \in K$ is defined as $v_{i,k}$. A vehicle $v$ consists of dimensions ${l_{v,i}, w_{v,i}}$ where $l_{v,i}, w_{v,i}$ are the length and width of the vehicle respectively. An example of a four legged intersection and vehicles arriving towards the intersection are depicted in Figure \ref{fig:Intersection}.}

\textbf{Intersection region and starting point:} Intersection region is defined as the common intersecting area of all road segments at the intersection ($I$). We define the point where a road meets the intersection region as the \emph{intersection starting point}. The intersection region and intersection starting point are shown Figure \ref{fig:Intersection}.

\textbf{Control region:} The control region of a road segment $k$ is a road segment of length $L$ measured from the intersection starting point along road segment $k$, where $L$ is defined as the control region length. The control region of the intersection consists of control regions of all road segments. 

\textbf{Vehicle position:} The position of vehicle $v_{i,k}$, $x_{v_{i,k}}(t)$ is the distance from the intersection starting point to the vehicle's front bumper at time $t$.


\textbf{Safety:} Safety at the intersection and control regions at time $t$ is achieved when any two vehicles' occupied areas do not have overlap, where the occupied area of a vehicle is a 2-d space covered by $l_{v,i} \times  w_{v,i}$ dimensions of the vehicle w.r.t. its current position.

\textbf{Travel time:} The travel time of a vehicle $v_{i,k}$, $TT_{v_{i,k}}$ is defined as the time taken for the vehicle to cross the intersection, i.e. the time between when the vehicle enters and exits the control region.

\textbf{Total travel cost:} Given that there are $N_{t}$ vehicles inside the control region at time $t$, the total travel cost $TTC(t)$ is the sum of the expected travel times of all $N_{t}$ vehicles.

\textbf{Problem Statement:} Given that the vehicle arrival times to control region are stochastic (the vehicle arrival times are not \emph{known a priori}), our objective is to  minimize the $TTC(t)$ at each time step while guaranteeing safety at every time step. 

\section{Intelligent Autonomous Intersection Management} \label{sec-polling-system}

A solution to the aforementioned optimization problem requires every vehicle in the control region to have a distinct scheduled time to reach the intersection starting point, and a well-defined trajectory to reach the intersection starting point at the scheduled time. \textcolor{darkgreen}{We can reduce the computational complexity of the overall optimization problem by decoupling these two optimizations as two separate sets of agents, which can act cooperatively. Two optimization problems are (a) a scheduling optimization problem and (b) a trajectory optimization problem. Thus, we propose a multi-agent solution named \textbf{Intelligent Autonomous Intersection Management (I-AIM)} which consists of two main components. The first component is a \emph{Polling-based Coordinating} Agent that schedules a time to arrive at the intersection region for every vehicle in the control region (scheduling optimization). The second component is a set of distributed \emph{RL-based Trajectory Control Agents} (we will use the name RL-based Agent for short) each of which controls the trajectory of a vehicle so that the vehicle reaches the intersection region precisely at the scheduled time given by the coordinating agent at the maximum speed (trajectory optimization)~\cite{max-speed}.} In the next sections we describe the two types of agents in detail. 

\vspace{-0.3em}
\subsection{Polling-based Coordinating Agent} \label{sec-polling-system}
\textcolor{darkgreen}{The Polling-based Coordinating Agent uses a polling algorithm and communicates with the RL-based Agent to send/receive relevant information. We first show how a vehicle's arrival can be modelled as a polling system. We then formulate the polling system for an intersection with single-lane roads, similar to \cite{polling-system}. Finally, we extend the formulation to a multi-lane intersection with multiple turnings.}

Each incoming road segment can be represented as a queue, and elements in the queue (customers) are the vehicles within the control region. The \emph{service time} should ensure that two vehicles from the same road will not reach the intersection simultaneously. Thus, the \emph{service time} is set to the time duration between when  the vehicle's front bumper enters the intersection region, and the vehicle's rear bumper enters the intersection region while travelling at the maximum speed. The \emph{switch over time} should ensure that a vehicle will enter the intersection only when a vehicle from a different road segment has exited the intersection region. Thus, the \emph{switch over time} is set to the time duration between when the vehicle's front bumper enters the intersection region, and its rear bumper exits the intersection region while travelling at the maximum speed. 

Once the \emph{service} and \emph{switch over} times are defined, any regular polling policy \cite{polling-survey} can be used to compute a schedule for every vehicle in the control region, and all the time-schedules will be non-conflicting. The polling algorithm is event-triggered, implying that when a new vehicle enters the control region, a new schedule will be computed. Thus, the polling algorithm can work with vehicles that are arriving stochastically.  

The above formulation that represents an intersection as a polling system is capable of handling intersections with single-lane road segments only. When there are multiple lanes per road segment, there is more than one conflicting point (see Figure~\ref{fig:crossing-points}). To address this limitation in a computationally efficient manner, we propose a Multi-Lane Polling System discussed next.

\textbf{Multi-Lane Polling System: } The proposed system represents each lane in the intersection as a separate FIFO queue, as shown in Figure \ref{fig:crossing-points}. Each queue contains vehicles that have not yet finalized the time schedule. We denote all queues as a set $Q$. With multiple conflict points between queues in $Q$, the time that a queue has to wait after processing another queue is not constant across queues compared to the single-lane scenario. Thus, instead of using a fixed service or switch over times we propose a queue dependent function named \emph{queue transition function} $f_q:q_1 \times q_2 \mapsto \mathbb{R}$. The function represents the time that the polling system should wait to process a customer (vehicle) in $q_2$ after servicing a customer from $q_1$, i.e. it represents the switch over time between $q_1$ and $q_2$ plus the service time of $q_1$. For two consecutive requests from the same queue, the $f_q$ output equals to the queue's service time. For two queues in the same road, $f_q$ is equal to zero as queues can be processed in parallel. For the scenario shown in Figure~\ref{fig:crossing-points} the queue transition function is a 4x4 matrix. Once the $f_q$ is defined, the polling order can be computed using the following algorithm.
\begin{algorithm}
    \KwIn{$Q=\{q_1, ... q_N\}$:$N$ number of queues for each lane}
    \KwIn{$\tau$ : time to reach the intersection at max speed from control region length $L$}
    
    $T_a = \{t_1, ... t_N \}$ : \text{latest service time of each queue initialized with zeros}
    
    $q \gets \text{select the queue with the first arrived vehicle}$
    
    \While{queues in Q are not empty}{
        $v \gets q.pop()$ \\
        $t_v \gets \tau + \max_{q_i}(T_a[q_i] + f_q(q,q_i))$ : \text{ignore queues with} $f_q(q,q_i))$ equals to zero  \\
        $T_a[q] \gets t_v$ \\
        $q \gets \text{select the next queue to process}$ \\
    }
    \caption{Multi-Lane Polling Algorithm}
    \label{algo-polling}
\end{algorithm}

With a given $Q$, the algorithm first initializes the set $T_a$ which is used to store the latest scheduled time (of the vehicle) assigned to each queue~\footnote{\textcolor{darkgreen}{The vehicles are allowed to change lanes (i.e. change the queue) inside the control region, until a vehicles receives a finalized time schedule.}}. In line 2, the first queue to process is selected. In line 4, the algorithm takes the first vehicle $v$ from the queue $q$ and finds the latest safest schedule time $t_v$ that can be assigned to $v$. Taking $\max_{q_i}(T_a[q_i] + f_s(q,q_i))$ ensures that the assigned time will not conflict with previously assigned times for all queues. Then, $t_v$ is stored in $T_n$ as the latest scheduled time for $q$. The next queue to process will be selected using the aforementioned polling policies (e.g. exhaustive, k-limited, or gated) in line 7. The only added computation over the standard polling algorithm is the $max$ operation (linear in the number of queues).

\textcolor{darkgreen}{\textbf{Extending to all roads segments and multiple turning directions: } For simplicity, we explained our Polling-based Coordinating Agent only with two road segments with vehicles moving straight through the intersection. Extending this setup to all road segments is trivial. As in Figure \ref{fig:Intersection}, when we have eight incoming lanes connecting to the same intersection, the queue transition function then becomes an 8x8 matrix. Then, to handle right and left turns, let us assume that each lane can have two allowed directions (either straight and left turn or straight and right turn). Then, each lane can be represented by two distinct queues, and the total number of queues will be 16. Finally, the queue transition function becomes a $16 \times 16$ matrix. We can generalize this matrix to $(n_l*2) \times (n_l*2)$, where $n_l$ is the number of incoming lanes. This matrix then can be used in \textbf{Multi-Lane Polling Algorithm} (Line 5).  Note that the matrix size does not increase exponentially in real-world intersections as most intersections consists of 4 or 5 incoming roads. The lanes per road segment are within the range of 2 to 5 incoming lanes.}

\subsection{RL-based Trajectory Control Agent} \label{sec-rl-based-tc}

Once the Polling-based Coordinating Agent schedules a time for a vehicle to reach the intersection region, the next objective is to control the vehicle's speed so that the vehicle reaches the intersection region precisely at the scheduled time. A vehicle needs to reach the intersection region at maximum speed to ensure the \emph{service} and the \emph{switch over} times are not violated. The higher the speed the higher the throughput will be \cite{max_speed}. 
A naive way to achieve this arrival to the intersection at a given time and at maximum speed is by stopping at a considerable distance away from the intersection starting point for some time and then only accelerating at the last minute. This sort of trajectory will block vehicles that follow, which reduces the control region's capacity \cite{polling-system}. Thus, an optimum trajectory should stay as close as possible to the intersection. In previous work \cite{polling-system}, this is achieved by solving a linear program for every vehicle starting from the leading vehicle in a road in a sequential order. However, solving a linear program every time the Polling-based Coordinating Agent updates the time schedules requires a significant amount of computation. If the trajectory is not computed fast enough, there is a risk of crashing with a former 
 vehicle or not reaching the intersection at a scheduled time. Furthermore, LP-based methods require the complete trajectory of the front vehicle to be known when computing the current vehicle's trajectory. This information is however unavailable when the front vehicle is not a CAV. Even if the front vehicle is a CAV, a re-computation is needed if the front vehicle's trajectory is changed. To overcome these limitations, we propose a Deep RL-based distributed solution where an agent is trained to control the vehicle's speed to reach the intersection precisely at the scheduled time. Once an RL-based Agent is trained offline, it can be used to make real-time decisions. This approach mitigates the safety concerns posed by the computational overhead of LP.    

We use the Q-learning algorithm to learn the vehicle's trajectory 
to reach the intersection region at a maximum speed (a.k.a. trajectory control). While learning to control the vehicles' trajectory to reach the intersection region, the agent should learn to keep a safe distance from the former vehicle (a.k.a. cruise control). See Section \ref{Trajectory control task} for the state, action, reward definitions. 
Combining these two objectives can be achieved by combining two reward functions from each task \cite{Mo-MDP}. However, we observe that there is a unique property of this problem that makes a standard Q-learning algorithm unsuitable. In the trajectory control task, successfully reaching the intersection is determined only when the vehicle reaches the intersection starting point (i.e. at the end of an episode), making the objective long term. Whereas in the cruise control task, the former vehicle can be humanoid/autonomous, and its behavior can be changed in short time intervals, i.e. the vehicle behaviour is only deterministic in a short time interval. The existing Q-learning algorithm can be set either as a short-term or as a long-term objective task using the \emph{discount factor}. Suppose the discount factor is set as a long-term task, the former vehicle's behavior cannot be adequately modeled and will result in a sub-optimal behavior (i.e. keeping a large distance from the former vehicle). If the discount factor is set as a short-term task, the RL-based Agent is unable to see a large number of time-steps ahead. Then the trajectory control task cannot be achieved properly because it requires controlling the vehicle a long distance away from the intersection. We propose a novel Q-learning algorithm coined \textbf{Multi-discount Q-learning} to learn the two tasks to preserve the short-term goal while following the long-term goal.

\section{Learning Long-term and Short-term Goals}

This section first introduces a Multi-objective MDP (MO-MDP) and explains why the standard Q-learning cannot learn multiple tasks with different temporal objectives. MO-MDP is different from MDP only in the reward function definition. In MO-MDP the reward function is a vector $\mathbf{r_t} \in \mathbb{R}^k$ where $k$ is the number of objectives/tasks. Learning with MO-MDP be considered for 2 main use cases, based on the information available about the MDP. The first scenario is when we know how much weight should be given to each objective (relative importance of each reward/task). That means we know a weight vector $\mathbf{w} \in \mathbb{R}^k$, so that we can compute $\mathbf{w.r_{t}}$. The second scenario is when we do not know how much weight should be given to each objective while learning, i.e., $\mathbf{w} $ is unknown. The former is solved by transferring into a single-objective MDP, and the latter is solved by learning objectives separately and combining them at the decision-making stage~\cite{Mo-MDP}.

Our problem consisting of the trajectory control and  cruise control tasks belongs to the former case where we know the weight vector $\mathbf{w}$. However, in our case, each task contains a different temporal objective. Before introducing the proposed algorithm, let us discuss why traditional Q-learning is unsuitable for this problem. First, we can modify the standard Q-learning equation (Equation \ref{n-step-q-learning}) to work with weight vector $\mathbf{w}$ in the MO-MDP defined above as follows.

\begin{equation}
\label{mo-n-step-q-learning}
    Q(s_t,a) \gets Q(s_t,a_t) + \alpha[ \sum_{\tau=0}^{n-1} \gamma^{\tau}(\mathbf{ w.r_{t+\tau}})  + \gamma^n \max_{a^\prime} Q(s_{t+n}, a^\prime) - Q(s_t,a_t)]
\end{equation}

Note that in Equation \ref{mo-n-step-q-learning}, when combining $k$ number of tasks, discount factor $\gamma$ is a scalar value between 0-1 and it is a common value for all the elements in $\mathbf{r_t}$. The discount factor determines how much importance is given to future rewards at the current time step $t$. The lower values of discount factor (0-0.9) typically have a short-term effect because the amount of future rewards considered can be approximated as $1/(1-\gamma)$ \cite{sutton}. For example, in the extreme case when $\gamma$ is 0, we only consider the immediate reward, while with $\gamma$ of 0.9 only the next 10 rewards have an impact. In both cases, the value of $\max_{a^n} Q(s^n, a^n)$ and rewards from $t+1$ to $t+n$ have less impact to the current value. The objective we are trying to achieve becomes a short-term objective. When $\gamma$ is 1, all the future values of reward have the same importance. In that case, the objective becomes a long-term objective. Since we can only assign a single scalar value to the discount factor $\gamma$, the combined task becomes either a long-term objective task or a short-term objective task. 

\textbf{Multi-discount Q-learning:} We introduce two components to overcome the aforementioned limitation. First, we introduce a discount vector $\boldsymbol\Gamma \in \mathbb{R}^k$ which can assign different discount factors to each objective for $k$ objectives of Multi-objective MDP. The expected return defined in Section \ref{sec-background} can be modified as $G_t = \sum^T_{\tau=t}\sum_{i=0}^k {\gamma_i^{\tau-t}r_{t+\tau,i}}$, where $\gamma_i \in \boldsymbol\Gamma$ represents the discount factor assigned to the $i^{th}$ objective and $r_{\tau,i}$ is the reward received for the $i^{th}$ objective at time step $t+\tau$. Each reward is associated with its own discount factor rather than a common one for all rewards. Since each reward is discounted differently, considering the reward is long-term or short term, temporal length of each objective is preserved.

The second aspect considers how to discount Q-value in Equation \ref{mo-n-step-q-learning}. The term $\max_{a^\prime} Q(s_n, a^\prime)$ in Equation \ref{mo-n-step-q-learning} is a scalar value and we cannot use discount vector $\mathbf\Gamma$ directly for the discounting. To preserve the temporally different objectives, we introduce a function to determine the discount factor based on the obtained reward vector. The function is named \emph{reward dependent discount function} $f: \mathbb{R}^k \mapsto [0,1]$. Using $f$, we can preserve the temporally different objectives because the discount value can be changed. 
Finally, we can write the multi-discount Q-learning equation as follows.

\begin{equation}
 \label{mo-discount-n-step-q-learning}
    Q(s_t,a_t) \gets Q(s_t,a_t) + \alpha[ \sum^{n-1}_{\tau=0}\sum_{i=0}^k {\gamma_i^{\tau}r_{t+\tau,i}} + f\mathbf(\mathbf{r_{t+\tau}})^n\max_{a^\prime} Q(s_{t+n}, a^\prime) - Q(s_t,a_t)]
\end{equation}

\textbf{Multi-discount Deep Q-learning (MD-DQN):} Deriving the Multi-discount Deep Q-learning from Equation \ref{mo-discount-n-step-q-learning} is straightforward. We can define the \emph{target function} in Equation \ref{target} for MD-DQN with \emph{n-step} return as show in Equation \ref{mo-target}. Then, this can be used to compute the loss function in Equation \ref{deep-q-learning} by replacing $y_{t}^{DQN}$ with $y_{t}^{mo}$.
\vspace{-0.5em}
\begin{equation} \label{mo-target}
    y_{t}^{mo}=\sum^{n-1}_{\tau=0}\sum_{i=0}^k {\gamma_i^{\tau}r_{t+\tau,i}} + f\mathbf(\mathbf{r_{t+\tau}})^n \max_{a^\prime} Q(s_{t+n}, a^\prime; \theta_t^-)
\end{equation}

\subsection{Trajectory Optimization as an MDP} \label{Trajectory control task}
We now formulate the trajectory control and cruise control problem as an MDP. For each task there is a separate reward.

\textbf{States:} The state space consists of six states. First three states are related to trajectory control and the rest are related to cruise control. The first state is the current speed of the vehicle. The second is the distance to the intersection region. The third is the remaining time to reach the intersection region. The fourth is the front vehicles speed. The fifth is the distance between the vehicle and the front vehicle. The last one is the acceleration of the front vehicle.

\textbf{Actions:} The action space consists of three actions; 1) brake action, 2) gas action, and 3) no-op action. The actual values assigned to these actions are -1, 0, and 1, respectively. A similar set of actions were used for vehicle speed control \cite{Cruise_control}. %

\textbf{Reward vector:} The reward vector in this problem is a 2-d vector $\mathbf{r} = \{r_1, r_2\}$ where $r_1$ and $r_2$ are the rewards received for the cruise and trajectory control tasks respectively. Note that all the numerical values used below for rewards were selected experimentally to balance both tasks.

In the trajectory control task, the following reward is given at the end of an episode based on whether the vehicle was able to reach the intersection starting point at the scheduled time.

\begin{equation}
\nonumber
    r_{1,end}=
    \begin{cases}
      10 + 3*s_v(T_s), & \text{reached the intersection at }T_s \\
     -10 & \text{otherwise}
    \end{cases}
  \end{equation}

where $T_s$ is the scheduled time and $s_v(T_s)$ is the speed of the vehicle at the end of the episode. This reward encourages the vehicle to reach the intersection exactly at the scheduled time at high speed. 

Besides $r_{1,end}$ at the end of an episode, at every time step $t$, $r_{1,step}$ is given as the negative value of the vehicle position normalized by the control region length.

\begin{equation} \label{eqn-step-reward}
    r_{1,step} = -x_{v}(t)/L \quad \text{at every time step}
\end{equation}

This encourages the vehicle to stay closer to the intersection and allows vehicles that follow enough space to enter the control region. 

In the cruise control task, $r_2$ is assigned based on the gap between the front vehicle. This reward encourages the vehicle to keep a safe gap whenever possible. A large negative reward is given to avoid crashing with the front vehicle.

\begin{equation}
\nonumber
    r_2=
    \begin{cases}
       -400 & \text{if a vehicle crashed with the front vehicle} \\
       0.1 & \text{else if $6m < gap < 20m$ } \\
       -0.1 & \text{else if $ gap < 6m$ } \\
     0 & \text{otherwise}
    \end{cases}
 \end{equation}
 
\textbf{Multi-discount Q-learning for trajectory control: } We formulate MD-DQN for the AIM problem as follows. We denote discount vector $\boldsymbol\Gamma$ as $\{d_1, d_2\}$ and the reward dependent discount function, $f(\textbf{r})$ defined as $d_2$ when $r_2$ is non-zero and $d_1$ when $r_2$ is zero \footnote{Here, $d_1, d_2$ are selected as 0.9 and 1 through a parameter sweep}.

\section{Experimental Methodology}
\label{sec-exp-meth}
We conducted three types of experiments. The first set of experiments evaluate the MD-DQN against Q-learning. The next set of experiments evaluate the trajectory computed by MD-DQN compared to an optimal policy computed using linear programming. \textcolor{new}{Since MD-DQN computes the trajectory, these experiments are designed to compare the trajectory computed by MD-DQN with other baselines and to show the trajectory is near-optimal. The last set of experiments is designed to evaluate the overall impact of I-AIM in an intersection compared to other baselines in terms of travel time and safety.} We conduct the above experiments with a microscopic traffic simulator, SMARTS~\cite{Ramamohanarao2016} using two experimental setups discussed next. The first setup is used to evaluate the MD-DQN Agent in the first and second experiments mentioned above, and the second setup is used for the third experiment mentioned above.

\textbf{Vehicle Experimental Setup (VE-setup)}: This  setup includes two vehicles (\emph{leading vehicle} and \emph{following vehicle}) that are arriving to an intersection on the same road. The \emph{following vehicle} is given a time (chosen uniformly at random) to reach the intersection, and is controlled by the RL-based Agent. The \emph{leading vehicle} can be either an autonomous or a human-driven vehicle (in which the RL-based Agent does not control the behavior).

\textbf{Intersection Experimental Setup (IE-setup)}: This experimental setup includes a 4-legged intersection with 2-lane roads, and vehicle arrivals are modeled as a Poisson distribution. Whenever a new vehicle arrives, the polling system computes a time-schedule for every vehicle, and then, RL-based Agents per-vehicle are used to control the vehicle speed.

\textcolor{darkgreen}{\textbf{Baselines}:
First, to compare the effectiveness of multi-objective learning of the RL-based Agent, we use threshold lexicographic DQN (TLDQN) proposed by Li \emph{et al.}~\cite{aamas_mqdqn}. We represent the trajectory and cruise control tasks as two separate Q-networks with lexicographic orders. At every time step, possible actions with higher Q-values are selected from each Q-network, and then the final action is chosen by considering the lexicographic order.}

\textcolor{darkgreen}{For the trajectory optimization task, we use two baselines. First, we use the same linear program formulation adopted by Miculescu \emph{et. al}~\cite{polling-system} for the trajectory control task to find the optimal solution. We denote this baseline as \textbf{LP-AIM}. Second, we use a heuristic solution for the trajectory optimization task, similar to the one proposed by Au \emph{et al.}~\cite{AIM_analytical}. In Au \emph{et al.} the objective is to optimize the time to reach the intersection. In contrast, in our trajectory optimization task, we maximize the end velocity and minimize the value in Equation \ref{eqn-step-reward}. We denote this baseline as \textbf{H-AIM}.} 

\textcolor{darkgreen}{For the schedule optimization task (the coordinating agent), we use traditional First-Come-First-Serve scheduling~\cite{AIM_definition} as a baseline. We denote this baseline as \textbf{FCFS-AIM}.}

\subsection{Evaluation Metrics}

We evaluate our solution based on the following metrics.

\textbf{Trajectory Performance:} The quality of trajectory $trj$, can be measured by computing how close the vehicle was to the intersection throughout the trajectory, leaving enough space for the vehicles behind. 
This can be computed as $X(trj) = \sum^{T_s}_{t=0} x_v(t)_{trj}/L$ where $x_v(t)_{trj}$ is the position of the vehicle at time $t$. The lower the  $X(trj)$ value \textcolor{new}{indicates the vehicle has stayed closer to the intersection, which is better as described in Section \ref{sec-rl-based-tc}}. Equation \ref{eqn-trj-comp} can be used to compare two trajectories ($trj_1$ and $trj_2$) by computing the difference.
\vspace{-0.3em}
\begin{equation} \label{eqn-trj-comp}
diff(trj_1, trj_2) = |X(trj_1) - X(trj_2)|
\end{equation}

Because the travel time depends only on the Polling-based Coordinating Agent, the travel time is an unsuitable metric to compare LP and MD-DQN in \textbf{VE-setup}. Thus, the above metric is used to compare the LP and the MD-DQN trajectory.

\textbf{Total Average Travel Time:} The total travel cost defined in Section \ref{sec-prob-def} can be used to calculate the travel time reduction for the entire simulation period and used in \textbf{IE-setup}.

\subsection{Parameter Settings}
The following parameters have been set: max. speed is $80kmh$, max. acceleration/deceleration is $2ms^{-2}$ and the discrete simulation step size $\Delta t$ is $0.2s$ which are the default values of SMARTS. \textcolor{new}{The control length $L$ is set to 400m, similar to the previous work \cite{polling-system}. The minimum value of \emph{service} and \emph{switch over} times in the queue transition function are set to 1 second so that the vehicles have enough time to cross the conflicting squares (See Figure \ref{fig:crossing-points}) at the maximum speed.} For vehicles with higher length ($l_{v,i}$), \emph{service} time is however increased to allow a safe traversal.

While training the RL-based Agent~(using \cite{acme}), the learning rate is set to 1e-5, exploration factor $\epsilon$ is annealed from 1 to 0 over 120000-time steps. The discount factor $\gamma$ is in the range of 0.9-1. \textcolor{new}{The RL parameters are selected through a parameter sweep.} The primal-dual gap of the Gurobi solver
is set to 1-e5 to avoid high computation times that occur when finding an optimal solution. The experiments are conducted in a server with an Intel Xeon(R) processor, 24 GB RAM, and an Nvidia GRID P40 GPU.

\pgfplotstableread[col sep = comma]{figures/reward_11.csv}\rewarddata
\pgfplotstableread[col sep = comma]{figures/reward_12.csv}\rewarddatatldqn
\pgfplotstableread[col sep = comma]{figures/reward_15.csv}\rewarddatamd
\pgfplotstableread[col sep = comma]{figures/reward_16.csv}\rewarddatanine

\pgfplotstableread[col sep = comma]{figures/reward_1_11.csv}\rewardonedata
\pgfplotstableread[col sep = comma]{figures/reward_1_12.csv}\rewardonedatatldqn
\pgfplotstableread[col sep = comma]{figures/reward_1_15.csv}\rewardonedatamd
\pgfplotstableread[col sep = comma]{figures/reward_1_16.csv}\rewardonedatanine

\pgfplotstableread[col sep = comma]{figures/reward_2_11.csv}\rewardtwodata
\pgfplotstableread[col sep = comma]{figures/reward_2_12.csv}\rewardtwodatatldqn
\pgfplotstableread[col sep = comma]{figures/reward_2_15.csv}\rewardtwodatamd
\pgfplotstableread[col sep = comma]{figures/reward_2_16.csv}\rewardtwodatanine

\begin{figure*}[]

\begin{subfigure}{0.32\columnwidth}
\begin{tikzpicture}
\begin{axis}[
xtick={100000, 200000, 300000, 400000, 500000},
height=4.8cm,
width=5.5cm,
xmax=360000,
label style={font=\footnotesize},
ylabel near ticks,
ylabel={$r_1 + r_2$},
xlabel={steps},
legend style={
legend cell align={left},
legend pos=south east,
legend columns=2,
font=\tiny,
legend image post style={xscale=0.5}
}
]
\addplot[blue, line width=0.3mm] table [mark=none, x=Step, y=Smoothed Value, col sep=comma]{\rewarddata};
\addplot[red, line width=0.3mm] table [mark=none, x=Step, y=Smoothed Value, col sep=comma]{\rewarddatanine};
\addplot[green1, line width=0.3mm] table [mark=none, x=Step, y=Smoothed Value, col sep=comma]{\rewarddatatldqn};
\addplot[brown, line width=0.3mm] table [mark=none, x=Step, y=Smoothed Value, col sep=comma]{\rewarddatamd};

\legend{DQN-1, DQN-0.9, TLDQN, MD-DQN}
\end{axis}
\end{tikzpicture}

     \caption{Total episodic reward: ($r_1 + r_2$)}
     \label{fig:total reward}

\end{subfigure}\hspace{5mm}\begin{subfigure}{.32\columnwidth}
\begin{tikzpicture}
\begin{axis}[
xtick={100000, 200000, 300000, 400000, 500000},
height=4.8cm,
width=5.5cm,
xmax=360000,
label style={font=\small},
ylabel near ticks,
ylabel={$r_{1,end}$},
xlabel={steps},
legend style={
legend cell align={left},
legend pos=south east,
font=\scriptsize,
}]
\addplot[blue, line width=0.3mm] table [mark=none, x=Step, y=Smoothed Value, col sep=comma]{\rewardonedata};
\addplot[red, line width=0.3mm] table [mark=none, x=Step, y=Smoothed Value, col sep=comma]{\rewardonedatanine};
\addplot[green1, line width=0.3mm] table [mark=none, x=Step, y=Smoothed Value, col sep=comma]{\rewardonedatatldqn};
\addplot[brown, line width=0.3mm] table [mark=none, x=Step, y=Smoothed Value, col sep=comma]{\rewardonedatamd};

\end{axis}
\end{tikzpicture}

     \caption{Trajectory control reward}

     \label{fig:trajectory reward}
\end{subfigure}\hspace{5mm}\begin{subfigure}{0.32\columnwidth}
\begin{tikzpicture}
\begin{axis}[
xtick={100000, 200000, 300000, 400000, 500000},
height=4.8cm,
width=5.5cm,
xmax=360000,
label style={font=\small},
ylabel near ticks,
ylabel={$r_2$},
xlabel={steps},
legend style={
legend cell align={left},
legend pos=south east,
font=\scriptsize
}
]
\addplot[blue, line width=0.3mm] table [mark=none, x=Step, y=Smoothed Value, col sep=comma]{\rewardtwodata};
\addplot[red, line width=0.3mm] table [mark=none, x=Step, y=Smoothed Value, col sep=comma]{\rewardtwodatanine};
\addplot[green1, line width=0.3mm] table [mark=none, x=Step, y=Smoothed Value, col sep=comma]{\rewardtwodatatldqn};
\addplot[brown, line width=0.3mm] table [mark=none, x=Step, y=Smoothed Value, col sep=comma]{\rewardtwodatamd};

\end{axis}
\end{tikzpicture}

     \caption{Cruise control reward ($r_2$)}
     \label{fig:cruise reward}

\end{subfigure}

\caption{The reward (moving average with window size of 10) achieved with fixed discount vs multi-discount Q-learning.}
\label{exp-mulit-discount}

\end{figure*}
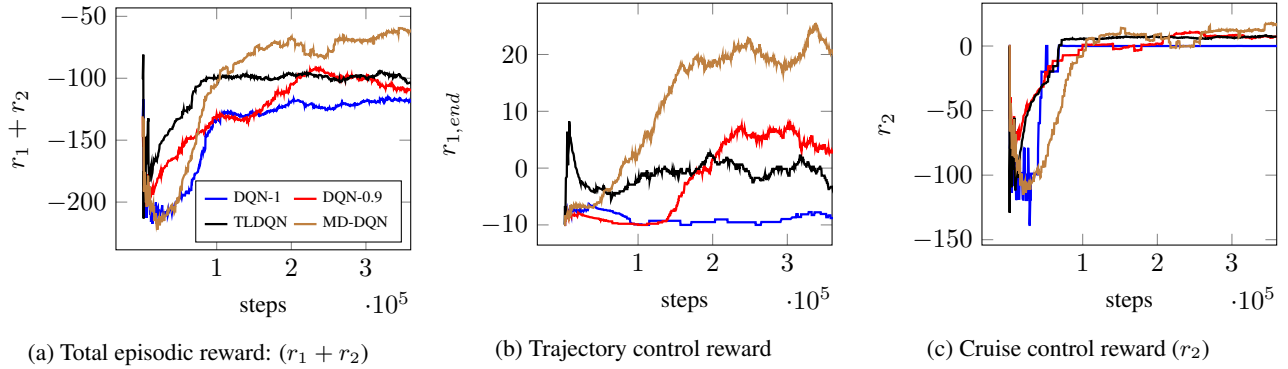
\pgfplotstableread[col sep = comma]{figures/lp_comparison.csv}\lpdata

\begin{figure}[t]
\begin{subfigure}{0.47\columnwidth}
\begin{tikzpicture}
\begin{axis}[
height=4.2cm,
width=6cm,
xmax=32,
xlabel near ticks,
label style={font=\footnotesize},
ylabel={Intersection distance},
xlabel={Scheduled time (s)},
legend columns=2,
legend style={at={(0.2,0.5)},anchor=west,font=\fontsize{7}{2}\selectfont, legend image post style={xscale=0.5}}
]

\addplot[red, mark=*, mark size = 0.4pt] table [mark=*, x=index, y=Gurobi, col sep=comma]{\lpdata};
\addplot[brown, mark=*, mark size = 0.4pt] table [mark=*,x=index, y=diff, col sep=comma]{\lpdata};
\addplot[blue, mark=*, mark size = 0.4pt] table [mark=*, x=index, y=DQN, col sep=comma]{\lpdata};

\legend{LP, diff, MD-DQN}
\end{axis}
\end{tikzpicture}

\caption{Trajectory: LP vs MD-DQN}
\label{fig:lp-dqn diff}
\end{subfigure}\hspace{2mm}\begin{subfigure}{0.48\columnwidth}
\begin{tabular}{|c|c|c|}
\hline
     & \begin{tabular}[c]{@{}l@{}}MD- DQN\end{tabular}                      & LP   \\ \hline
avg. & 8.3e-4 & 0.65 \\ \hline
min. & 4.5e-4 & 0.03 \\ \hline
max  & 4.2e-3 & 3.37 \\ \hline
\end{tabular}
\centering
\caption{Execution times (s) for MD-DQN and LP action selection. The average, minimum and maximum are shown.}

\label{fig:lp-dqn time}
\end{subfigure}
\vspace{-0.6em}
\caption{Performance vs execution times for LP and MD-DQN}
\vspace{-1em}
\end{figure}

\section{Experimental Results} \label{sec-exp-results}

\textbf{Multi-discount Q-learning vs Q-learning:} This experiment evaluates the advantage of multi-discount Q-learning over traditional Q-learning. The experiments are conducted with the \textbf{VE-setup} described in Section \ref{sec-exp-meth}. The RL-based Agent is trained to control the \emph{following vehicle} to reach the intersection exactly at the scheduled time while avoiding collision with the \emph{leading vehicle}. During the training process, scheduled times are assigned uniformly at random at the start of an episode. 
To simulate short-term driving behavior, the \emph{leading vehicle} selects uniformly at random whether to accelerate, decelerate or maintain the speed every 2 secs. 

Figure \ref{exp-mulit-discount} shows the reward achieved over 360000 time-steps \textcolor{new}{(until the rewards are converged)} by each algorithm. Here, DQN-1 and DQN-0.9 refer to Q-learning with fixed discount factors of 1 and 0.9 respectively \textcolor{new}{which were selected from a parameter sweep}. Figure \ref{fig:total reward} shows the total episodic reward (combining rewards from the trajectory control and cruise control). There is a clear gain in the total reward of MD-DQN compared to fixed DQNs, and TLDQN which highlights the importance of MD-DQN \textcolor{new}{i.e. higher total reward means MD-DQN achieves both trajectory control and cruise control objectives better than other baselines.}

\textcolor{darkgreen}{Next, to analyze the agent behaviour further, we represent the reward in each sub-task using $r_{1,end}$ and $r_2$ in Figure \ref{fig:trajectory reward} and Figure \ref{fig:cruise reward} respectively. Let us first compare fixed DQNs with MD-DQN.} In Figure \ref{fig:cruise reward}, DQN-1 converges to 0 because with the discount factor 1, DQN-1 is unable to learn \textcolor{new}{how to follow the leading vehicle i.e. the short-term behavior of the \emph{leading vehicle}, and thus it keeps a large safe distance from the \emph{leading vehicle}.
Due to keeping a large distance, DQN-1 did not reach the intersection at the scheduled time which resulted in a lower gain for $r_{1,end}$ in trajectory control.}

In Figure \ref{fig:cruise reward}, DQN-0.9 achieves positive values as it can predict the \emph{leading vehicle's} behavior. However, in Figure \ref{fig:trajectory reward}, DQN-0.9 does not achieve a higher reward as it only sees a few time-steps ahead. Thus, DQN-0.9 does not have enough time to reach the intersection at maximum speed or exactly at the scheduled time. Therefore, the reward is less than what is achieved by MD-DQN. MD-DQN overcomes these limitations and achieves higher rewards in both Figure \ref{fig:cruise reward} and Figure \ref{fig:trajectory reward}.

\textcolor{darkgreen}{TLDQN is not able to achieve higher rewards than MD-DQN as shown in Figure \ref{exp-mulit-discount}. The performance of TLDQN is similar to DQN-0.9 in all the sub-figures. This is due to TLDQN learns tasks independently and only combines the
actions from each objective at the inference time, thus failing to learn the  inter-relationship between objectives. 
In contrast, MD-DQN is able to learn such relationships resulting in higher rewards.}

\textbf{Multi-discount Q-learning vs Optimal Control:} This experiment is focused on the optimality of multi-discount Q-learning. We use the same \textbf{VE-setup} used in the previous section and evaluate MD-DQN against linear programming (LP). As described in Section \ref{sec-rl-based-tc}, with LP the complete trajectory of the \emph{leading vehicle} should be available for the \emph{following vehicle} to compute its trajectory. Due to this reason, we can only conduct the experiments with the autonomous leading vehicle. For the comparison, we use the same LP formulation as in \cite{polling-system} and implemented using Gurobi. The pre-trained RL-based Agent from the previous section is used. 

Figure \ref{fig:lp-dqn diff} shows the trajectory performance of the trajectory computed by LP and MD-DQN. Both LP and MD-DQN are given the same scheduled times in the range of 20s to 32s \textcolor{new}{where 20s is the shortest possible time that a vehicle can traverse the control region length (400m) and 32s is the highest time scheduled by the intersection controller}. The resulting trajectory performance value $X$ is shown in the figure. Note that $X(\text{MD-DQN})$ achieves trajectory performance closer to the LP $X(LP)$ as the difference between trajectories, $diff(\text{MD-DQN}, LP)$ is less than 5 in majority of cases. \textcolor{new}{This means that throughout the entire trajectory in control region length $L$ (which is 400m), the total vehicle position difference is 5m, which is a small value compared to $L$.} Thus, MD-DQN is able to achieve near-optimal performance. As we show later, these minor deviations work in favor of the safety of the final solution.

Figure \ref{fig:lp-dqn time} shows the computational time required by LP and MD-DQN to select an action at different scheduled times. MD-DQN action selection is \textcolor{new}{two orders of magnitude faster, on average compared to that of LP}, as shown in the table. The maximum computation time of LP is 3.37 seconds. This is significant, since a vehicle traveling at 80 kmph may travel around 60m by the time the computation completes with LP, thus making the computed trajectory unsafe or unfeasible. \textcolor{new}{In MD-DQN vehicle will only travel 0.08m during the action selection, making the RL approach safe.}

This experiment highlights the benefit of MD-DQN in a real-time setting. With a small reduction in optimality, we are able to gain much lower computation time, which enables real-time AIM.  

\begin{table}[t]
\begin{tabular}{|l|l|l|l|}
\hline
Traffic level (\# vehicles)      & 530    & 1080   & 1750    \\ \hline
Dynamic Traffic Signal (DTS)(s)         & 77.59  & 235.73 & 543.18 \\ \hline
FCFS-AIM(s) & 15.21 & 131.46 & 388.57  \\ \hline
H-AIM(s) & 15.25 & 98.09 & 313.34  \\ \hline
LP-AIM(s) & 13.85 & 94.92 & 304.44  \\ \hline
I-AIM(s) & 13.66 & 92.76 & 302.63  \\ \hline
\end{tabular}
\centering
\caption{Total average travel time for dynamic traffic signal, LP-AIM, H-AIM and I-AIM for different traffic levels.}

\label{tb-travel-time}
\end{table}\begin{table}[t]
\begin{tabular}{|l|l|l|l|}
\hline
Traffic level (\# vehicles)      & 530    & 1080   & 1750    \\ \hline
H-AIM & 65 & 49 & 69  \\ \hline
LP-AIM & 1 & 66 & 51  \\ \hline
I-AIM & 0 & 0 & 0  \\ \hline

\end{tabular}
\centering
\caption{The number of vehicles that did not arrive to the intersection crossing point within 1 second from the scheduled time.}

\label{tb-deviations}
\end{table}

\textbf{I-AIM Evaluation:} This experiment focuses on the overall performance gain of I-AIM in reducing travel time, using the \textbf{IE-setup}. As a baseline, SMARTS's Dynamic Traffic Signals (DTS) are used.

\textcolor{darkgreen}{Table \ref{tb-travel-time} presents the total average travel time for I-AIM and dynamic traffic signal control with 30 minutes of simulation for different traffic levels (number of vehicles) \textcolor{new}{with a Poisson distribution as in Hu \emph{et al.}\cite{traffic-level}}. Regardless of the traffic level, AIM-based methods result in  substantially lower average travel time over DTS.
This is due to avoiding vehicles' stop-and-go nature and yellow signal time. I-AIM is substantially better than
the traditional FCFS-AIM (wherein vehicles traverse the intersection in the order of their arrival). 
This is because, with I-AIM, we are able to achieve platooning-like behavior, avoiding larger delays in switch over times while shifting from a road segment to road segment. 
H-AIM, LP-AIM (the trajectory computed by LP), and I-AIM result in similar travel times because these three methods use the same polling-based schedule controller proposed in Section \ref{sec-polling-system}, which has a crucial impact on the total travel time.
The slight differences in travel time between H-AIM, LP-AIM and I-AIM are due to the deviations in trajectories discussed next.}

\textcolor{darkgreen}{Even though these minor deviations
do not have a substantial impact on the travel time, they pose a significant safety risk. 
Table \ref{tb-deviations} shows the number of vehicles that did not arrive at the intersection crossing point within 1 second of their scheduled time. \textcolor{new}{Since the switch over time is 1 second, any vehicle that has deviated more than 1 second poses a safety risk of crashing with a vehicle coming from another lane.}
Both heuristic-based and LP methods optimize the trajectory based on an internal model or a set of linear equations. However, these models and equations may not necessarily match the simulation dynamics.
For instance, in LP methods, the equations used to compute the trajectory may not be aligned with the simulation environment precisely. \textcolor{new}{In the simulation environment, vehicles tend to have slight deviations from the intended path computed by LP due to changes in traffic.} Due to these reasons, heuristic and LP-based methods tend to have higher deviation percentages, violating the switch over time and thus posing a safety risk. In contrast, RL-based I-AIM achieves superior performance with no deviations, i.e. not a single vehicle violated the scheduled time throughout the entire simulation. 
This added safety level for RL-based Agents is attributed to the fact they are trained 
in the same simulation environment, allowing it to learn the vehicle and simulation dynamics. Further, since RL works iteratively, the RL-based Agent can adjust the vehicle acceleration even when vehicles are slightly deviating from the intended trajectory, thereby offering higher safety levels.}
\section{Conclusion}
\textcolor{darkgreen}{This paper presents a real-time deployable AIM solution that uses a Polling-based Coordinating Agent and a set of distributed RL-based Agents. The paper proposes a novel RL algorithm named \emph{multi-discount Q-learning} to effectively achieve the AIM task in a complex intersection. We demonstrate that MD-DQN can successfully learn  tasks with different temporal objectives, while the existing state-of-the-art approaches fail to do so}. 

\textcolor{new}{In the future, we plan to explore the coordination of I-AIM agents over multiple intersections. Finally, it would be interesting to explore the possibility of multi-discount Q-learning with opponent-based games where the opponent behavior is changing over time.}

\bibliographystyle{unsrt}  
\bibliography{references}

\end{document}


\maketitle

\section{Outline}

Supplementary material includes the following sections. The \textbf{Related Work} section provides a literature review of Autonomous Intersection Management from a traffic engineering perspective. The next section provides a brief introduction to Q-learning and proves the convergence of multi-discount Q-learning and necessary conditions for convergence. The final section provides further experimental details with one more baseline. 

\section{Related Work}

There are two inter-dependent sub-problems that have been studied in the literature to optimize the AIM; (1) find a distinct time-schedule for each incoming vehicle to arrive at the intersection, and (2) find a vehicle trajectory such that a vehicle arrives at the intersection at exactly at the schedule time. The existing work can be divided into two main categories based on how they solved aforementioned two problems.

The first category of work  focuses on finding a safe trajectory as a solution to sub-problem (2) \cite{AIM_2,AIM_1,AIM_4}, by employing a heuristic to compute the time-schedule to sub-problem (1). Even though these approaches are able to compute trajectories, they do not optimize the sub-problem (1). Thus, they do not optimize the throughput nor reduce waiting time at intersections. In contrast, our objective is to maximize the throughput at the intersection.

The second category of work optimizes the time-schedule (sub-problem (1)) as a scheduling problem and then computes vehicle trajectories which adhere to the optimized time-schedule \cite{AIM_5,AIM_6,AIM_7}. The scheduling optimization is in fact NP-hard \cite{np-hard}, which makes it computationally expensive. Another drawback of this kind of approach is that all the arrival times of vehicles need to be known \emph{beforehand} to optimize the time-schedule. Thus, such solutions cannot be applied in a real-time setting where vehicle arrival times are stochastic. 

Recent work \cite{polling-system} proposed a solution to optimize the throughput in a stochastic setting using a polling system and linear programming. The polling system is used to optimize the time-schedule for each vehicle. Then, a linear program is solved for every vehicle to find their trajectory. These linear programs need to be computed sequentially (centralized). That means the linear program for the first vehicle arriving at the intersection should be computed first, and then the next vehicle, and so on. Though this approach can successfully solve the stochastic AIM problem, computational time required for linear programming hinders the applicability of this solution to real-time. We propose a distributed learning-based solution to replace the centralized linear programming, which enables real-time AIM. 

\section{Convergence of Multi-discount Q-learning}

We prove the convergence of multi-discount Q-learning and its necessary conditions. We follow a similar method used by~\cite{Q-learning-proof, random_process} for the proof of convergence. 

\subsection{Q-learning} \label{sec-Q-learning}

In RL, first the problem needs to be formulated as a Markov-decision process (MDP). A MDP consists of a state space $\mathcal{S}$ and, an action space $\mathcal{A}$. When an action $a_t \in \mathcal{A}$ is taken  in the current state $s_t \in \mathcal{S}$, at time $t$, the MDP's state changes to $s_{t+1}$ according to the transition probability $\mathcal{T}(s_t,a_t,s_{t+1}) = Pr(s_{t+1}|s_t,a_t)$. The MDP also provides a reward  $r_t$ for the above transition where $r_t$ is assigned according to the function $\mathcal{R}(s_t,a_t,s_{t+1})$. A RL agent acting on the MDP consists of a policy $\pi(a|s)$ which describes the agent's behaviour. The policy $\pi(a|s)$ indicates the probability of agent taking action $a$ in state $s$. The objective of the agent is to maximize expected reward $G_t$ starting from any given time step $t$. The \emph{expected reward} is defined as $G_t = \sum^T_{\tau=t}{\gamma^{\tau-t}r_\tau}$, where $t$ is the current time step, $\gamma$ is the discount factor and $T$ is the time step that the MDP reaches a terminal state (also called an end of an episode). 

The action-value function (Q-function) for policy $\pi$ stores the expected reward by taking action $a$ in state $s$ (state-action pair) is defined as: $Q^{\pi}(s,a) = \mathop{\mathbb{E}}[G_t|s_t=s, a_t=a,\pi]$. The optimal Q-function is the Q-function with maximum values to all state-action pairs as defined by the recursive Bellman equation \cite{sutton}. 

\begin{equation}
    Q^*(s_t,a_t) = \sum_{s^\prime \in \mathcal{S}}\mathcal{T}(s_t,a_t,s^\prime)[r+\gamma \max_{a^\prime}Q^*(s^\prime, a^\prime)]
\end{equation}

where $a^\prime$ is the best action (action with the highest Q-values) that can be taken from state $s^\prime$. The optimal policy $\pi^*$ is the policy that selects actions with the highest Q-values for every state from the optimal Q-function.

Often the MDP's transition probability $\mathcal{T}$ is an unknown. Q-learning computes the optimal Q-function in an iterative manner by observing the transitions ($s_t, a_t, s_{t+1}, r_t$) at every time step. Considering the reward from n number of steps we get the following Q-learning equation.

\begin{equation} \label{n-step-q-learning}
    Q(s_t,a_t) \gets Q(s_t,a_t) + \alpha[ \sum_{\tau=0}^{n-1} \gamma^\tau r_{t+\tau} + \gamma^n \max_{a^\prime} Q(s_{t+n}, a^\prime) - Q(s_t,a_t)]
\end{equation}

\textbf{Deep Q-learning:} For the most real-world problems, number of state-action pairs grows exponentially. Thus we cannot store the Q-function in a tabular manner. Deep Q-learning (DQN) uses a neural network to approximate the tabular Q-function. The deep neural network is represented as $Q(s,a;\theta)$ where $\theta$ are the set of parameters of the neural network (Deep Q-network). DQN learns the Q-function by finding the $\theta$ parameters by minimising the following squared loss function at every time step $t$. 

\begin{equation} \label{deep-q-learning}
    L_t(\theta_t) = \mathbb{E}_{(s,a,r,s^\prime) \sim \mathcal{D}} [(y_t^{DQN} - Q(s,a;\theta_t))^2]
\end{equation}

with \emph{target function} $y_t^{DQN}$ being,

\begin{equation} \label{target}
    y_t^{DQN} = r + \gamma \max_{a^\prime} Q(s^\prime, a^\prime; \theta_t^-)
\end{equation}

where $\theta^-$ parameters represent a separate neural network called \emph{target network} that keeps its parameters ($\theta^-$) frozen for a number of time steps while ($\theta$) parameters are optimized. After a certain number of time steps $\theta$ parameters are copied to $\theta^-$. $\mathcal{D}$ represents a data-set built from agents' previous experiences which consists of tuples of $(s,a,r,s^\prime)$ which is known as \emph{experience replay}. Deep Q-network is trained by taking sampled mini-batches from $\mathcal{D}$ as in Equation \ref{deep-q-learning}.

\subsection{Convergence Proof}

We will prove the convergence for the 1-step multi-discount Q-learning and this can be easily extended for n-step  multi-discount Q-learning as well. We follow a similar proof used in \cite{Q-learning-proof}. 1-step multi-discount Q-learning can be written as follows. Here, we drop the time notation and represent current state as $s$ and next state as $s^\prime$. 

\begin{equation}
 \label{mo-discount-n-step-q-learning}
    Q(s,a) \gets Q(s,a) + \alpha[\sum_{i=0}^k {r_{i}} + f\mathbf(\mathbf{r})\max_{a^\prime} Q(s^\prime, a^\prime) - Q(s,a)]
\end{equation}

and the TD error is;

\begin{equation}
    \delta = \sum_{i=0}^k {r_{i}} + f\mathbf(\mathbf{r})\max_{a^\prime} Q(s^\prime, a^\prime) \\ - Q(s,a)
\end{equation}

For the optimal Q-function we can define the contraction operator $\mathcal{H}$ as follows.

\begin{equation}
    (\mathcal{H}Q)(s,a) = \sum_{s^{\prime} \in \mathcal{S}} \mathcal{T}(s,a,s^{\prime})[\sum_{i=0}^k {r_{i}} + f(\mathbf{r})\max_{a^\prime} Q(s^{\prime}, a^\prime)]
\end{equation}

\begin{lemma} \label{lemma}
Given the \emph{reward dependent discount function} $f: \mathbb{R}^k \mapsto [0,1]$ and two Q-functions $Q_1(s,a)$ and $Q_2(s,a)$.

\begin{equation}
    || \mathcal{H}Q_1 - \mathcal{H}Q_2||_{\infty} \leq \gamma^\prime ||Q_1 - Q_2||_\infty
\end{equation}

When $\textbf{r}$ is deterministic i.e. the transition $P(s^\prime|s,a)$ is 1 for some state $s_1 \in \mathcal{S}$ and zero elsewhere. 

where $||x||_\infty$ is the sup-norm of x and $\gamma^\prime$ is the upper-bound of the \emph{reward dependent discount function}. 
\end{lemma}

\begin{proof}
We follow similar step used in the proof in \cite{state-dependent}, which uses a state-dependent discount factor. 

\begin{align*}
    & || \mathcal{H}Q_1 - \mathcal{H}Q_2||_{\infty}  = \\
    &= \max_{s,a} |\sum_{s^{\prime} \in \mathcal{S}} \mathcal{T}(s,a,s^{\prime})[\sum_{i=0}^k {r_{i}} + f(\mathbf{r})\max_{a^\prime} Q_1(s^{\prime}, a^\prime)-\sum_{i=0}^k { r_{i}} - f(\mathbf{r})\max_{a^\prime} Q_2(s^{\prime}, a^\prime)]| \\
    & \text{Note that when transition are deterministic i.e. $P(s^\prime|s,a)$ is 1 for one state and zero elsewhere,} \\  & \text{$f(\textbf{r})$ becomes independent of $s^\prime$} \\
    &=\max_{s,a} f(\mathbf{r}) |\sum_{s^{\prime} \in \mathcal{S}} \mathcal{T}(s,a,s^{\prime}) [\max_{a^\prime} Q_1(s^{\prime}, a^\prime)- \max_{a^\prime} Q_2(s^{\prime}, a^\prime)]| \\
    & \text{Since } \exists \gamma^\prime \text{ s.t. } 0 \leq f(\mathbf(r)) \leq \gamma^\prime \leq 1 \\ 
    &= \max_{s,a} \gamma^\prime |\sum_{s^{\prime} \in \mathcal{S}} \mathcal{T}(s,a,s^{\prime})[\max_{a^\prime} Q_1(s^{\prime}, a^\prime)- \max_{a^\prime} Q_2(s^{\prime}, a^\prime)]| \leq \\
    &= \max_{s,a} \gamma^\prime \sum_{s^{\prime} \in \mathcal{S}} \mathcal{T}(s,a,s^{\prime})|\max_{a^\prime} Q_1(s^{\prime}, a^\prime)- \max_{a^\prime} Q_2(s^{\prime}, a^\prime)| \leq \\ 
    &= \max_{s,a} \gamma^\prime \sum_{s^{\prime} \in \mathcal{S}} \mathcal{T}(s,a,s^{\prime})|\max_{a^\prime} Q_1(s^{\prime}, a^\prime)- \max_{a^\prime} Q_2(s^{\prime}, a^\prime)| \\
    &= \max_{s,a} \gamma^\prime \sum_{s^{\prime} \in \mathcal{S}} \mathcal{T}(s,a,s^{\prime})\max_{x,y} |Q_1(x, y)- Q_2(x, y)| \\
    &= \max_{s,a} \gamma^\prime \sum_{s^{\prime} \in \mathcal{S}} \mathcal{T}(s,a,s^{\prime})||Q_1- Q_2||_\infty \\
    &= \gamma^\prime ||Q_1- Q_2||_\infty \\
\end{align*}

\end{proof}

Next, we use the following theorem used in \cite{random_process} to prove the convergence. 

\begin{theorem} \label{random process}
Given a iterative random process ($\Delta_t$) with n-dimensional vector as follows; the random process will converge to zero with probability 1;

\begin{equation}
    \Delta_{t+1}(x) = (1-\alpha_t(x))\Delta_t(x) + \alpha_t(x)F_t(x)
\end{equation}
\end{theorem}

when

\begin{itemize}
    \item $0 \leq \alpha_t \leq 1$, $\sum_t \alpha_t(x) = \infty$ and $\sum_t \alpha_t^2(x) < \infty $
    \item $||\mathop{\mathbb{E}}[F_t(x)|\mathcal{F}_t]||_W \leq \gamma||\Delta_t||_W$, with $\gamma < 1$ 
    \item $var[F_t(x)|\mathcal{F}_t] \leq C(1+||\Delta_t||_W^2)$, for $C > 0$
\end{itemize}

Here, $\mathcal{F}_t$ stands for all the history of $\Delta, F, \alpha$ until $t-1$ and $W$ is a weighted vector.

Now we prove the convergence of multi-discount Q-learning using the above theorem. 

\begin{theorem}

Given the Q-learning updates in a MDP;

\begin{equation}
 \label{mo-discount-n-step-q-learning}
    Q_{t+1}(s_t,a_t) \gets Q_t(s_t,a_t) + \alpha[\sum_{i=0}^k {r_{i,t}} + f\mathbf(\mathbf{r_t})\max_{a^\prime} Q(s^\prime_t, a^\prime) - Q(s_t,a_t)]
\end{equation}

Where we introduce $*_t$ term to represent the variables' value at current time step $t$.

The equation converges to optimal Q-values $Q^*(s,a)$ with probability 1 if

\begin{equation}
  \sum_t \alpha_t(s_t,a_t) = \infty   \text{ and  }   \sum_t \alpha_t^2(s_t,a_t) \leq \infty \text{ ;  } \forall (s_t,a_t) \in \mathcal{S} \times \mathcal{A}  
\end{equation}

\begin{proof}
The learning rate can satisfy the $\alpha_t(s_t,a_t)$ conditions by annealing the learning rate. This involves visiting every $(s,a)$ infinitely often.

Second, we can write the Q-learning equation as a random process by subtracting from optimal Q-value function as follows.

\begin{equation}
 \label{mo-discount-n-step-q-learning}
    Q_{t+1}(s_t,a_t) - Q^*(s_t,a_t) \gets (1-\alpha)Q_t(s_t,a_t) + \alpha[\sum_{i=0}^k {r_{i,t}} + f\mathbf(\mathbf{r_t})\max_{a^\prime} Q_t(s^\prime_t, a^\prime) - Q^*(s_t,a_t)]
\end{equation}

With analogy to random process;

\begin{itemize}
    \item $\Delta_t = Q_{t+1}(s_t,a_t)-Q^*(s_t,a_t)$
    \item $F_t(s,a) = \sum_{i=0}^k {r_{i,t}} + f\mathbf(\mathbf{r_t})\max_{a^\prime} Q_t(s^\prime_t, a^\prime) - Q^*(s_t,a_t)$
\end{itemize}

Then we need to prove that te condition 2 and 3 in Theorem \ref{random process} satisfied by Q-learning Equation. 

\begin{align*}
    \mathop{\mathbb{E}}[F_t(x)|\mathcal{F}_t] &= \sum_{s^{\prime} \in \mathcal{S}} \mathcal{T}(s,a,s^{\prime})[\sum_{i=0}^k { r_{i,t}} + f\mathbf(\mathbf{r_t})\max_{a^\prime} Q(s^\prime_t, a^\prime) - Q^*(s_t,a_t)] \\
    &=\mathcal{H}Q_t(s,a)-Q^*(s,a) \\
    \text{Since Q* is a fixed point} \\ 
    &=(\mathcal{H}Q_t(s,a)-\mathcal{H}Q^*(s,a) \\
    &\leq \gamma^\prime ||Q_t - Q^*||_\infty \; \text{  from Lemma \ref{lemma}} \\
    &\leq \gamma^\prime ||\Delta_t||_\infty
\end{align*}

This satisfies the Condition 2 in Theorem \ref{random process}.

\vspace{6cm}

Then we can write; 

\begin{align*}
    var[F_t(x)|\mathcal{F}_t] &= (\mathop{\mathbb{E}}[F_t^2(s,a)] - \mathop{\mathbb{E}}[F_t(s,a))]^2) \\
    &=\mathop{\mathbb{E}}[ ( \sum_{i=0}^k {r_{i,t}} + f\mathbf(\mathbf{r_t})\max_{a^\prime} Q(s^\prime_t, a^\prime) - Q^*(s_t,a_t) - \mathcal{H}Q_t(s,a)-Q^*(s,a))^2] \\
    &=\mathop{\mathbb{E}}[ ( \sum_{i=0}^k {r_{i,t}} + f\mathbf(\mathbf{r_t})\max_{a^\prime} Q(s^\prime_t, a^\prime) - \mathcal{H}Q_t(s,a)))^2] \\
    &= var[\sum_{i=0}^k {r_{i,t}} + f\mathbf(\mathbf{r_t})\max_{a^\prime} Q(s^\prime_t, a^\prime) | \mathcal{F}_t] \\
    & \text{when $\sum_{i=0}^k {r_{i,t}}$ is bounded whole term is bounded such that;} \\
    & \leq C(1+ ||\Delta_t||^2_W); \;\; \text{for some } C > 0
\end{align*}

This satisfies the Condition 3 in Theorem \ref{random process} and concludes the proof of convergence.

\end{proof}

\end{theorem}

\section{Results}

We show the same experiment results showed in our experimental data in the manuscript but added the DQN with a discount factor of 0.95 to provide a full view of the results. 

In these figures, DQN-0.95 achieves a higher reward in $r_2$ as it can see more time-steps ahead than the DQN-0.9. However, because the discount factor 0.95 is closer to a long-term objective, DQN-0.95 is not able to predict the \emph{leading vehicle} properly, which crashes with the \emph{front vehicle} leading to lower gain for $r_{1,end}$ in Figure \ref{fig:cruise reward}.

\pgfplotstableread[col sep = comma]{figures/reward_11.csv}\rewarddata
\pgfplotstableread[col sep = comma]{figures/reward_15.csv}\rewarddatamd
\pgfplotstableread[col sep = comma]{figures/reward_16.csv}\rewarddatanine
\pgfplotstableread[col sep = comma]{figures/reward_18.csv}\rewarddataninefive

\pgfplotstableread[col sep = comma]{figures/reward_1_11.csv}\rewardonedata
\pgfplotstableread[col sep = comma]{figures/reward_1_15.csv}\rewardonedatamd
\pgfplotstableread[col sep = comma]{figures/reward_1_16.csv}\rewardonedatanine
\pgfplotstableread[col sep = comma]{figures/reward_1_18.csv}\rewardonedataninefive

\pgfplotstableread[col sep = comma]{figures/reward_2_11.csv}\rewardtwodata
\pgfplotstableread[col sep = comma]{figures/reward_2_15.csv}\rewardtwodatamd
\pgfplotstableread[col sep = comma]{figures/reward_2_16.csv}\rewardtwodatanine
\pgfplotstableread[col sep = comma]{figures/reward_2_18.csv}\rewardtwodataninefive

\begin{figure*}[]
\begin{subfigure}{\columnwidth}
\begin{tikzpicture}
\begin{axis}[
xtick={100000, 200000, 300000, 400000, 500000},
height=6cm,
width=8cm,
xmax=360000,
label style={font=\footnotesize},
ylabel={$r_1 + r_2$},
xlabel={steps},
legend style={
legend cell align={left},
legend pos=south east,
font=\small,
legend image post style={xscale=0.5}
}
]
\addplot table [mark=none, x=Step, y=Smoothed Value, col sep=comma]{\rewarddata};
\addplot table [mark=none, x=Step, y=Smoothed Value, col sep=comma]{\rewarddatanine};
\addplot table [mark=none, x=Step, y=Smoothed Value, col sep=comma]{\rewarddataninefive};
\addplot table [mark=none, x=Step, y=Smoothed Value, col sep=comma]{\rewarddatamd};


\legend{DQN-1, DQN-0.9, DQN-0.95, MD-DQN}
\end{axis}
\end{tikzpicture}
     \centering
     \caption{Total episodic reward: ($r_1 + r_2$)}
     \label{fig:total reward}
\end{subfigure}

\begin{subfigure}{\columnwidth}
\begin{tikzpicture}
\begin{axis}[
xtick={100000, 200000, 300000, 400000, 500000},
height=6cm,
width=8cm,
xmax=360000,
label style={font=\small},
ylabel={$r_2$},
xlabel={steps},
legend style={
legend cell align={left},
legend pos=south east,
font=\scriptsize
}
]
\addplot table [mark=none, x=Step, y=Smoothed Value, col sep=comma]{\rewardtwodata};
\addplot table [mark=none, x=Step, y=Smoothed Value, col sep=comma]{\rewardtwodatanine};
\addplot table [mark=none, x=Step, y=Smoothed Value, col sep=comma]{\rewardtwodataninefive};
\addplot table [mark=none, x=Step, y=Smoothed Value, col sep=comma]{\rewardtwodatamd};


\end{axis}
\end{tikzpicture}
     \centering
     \caption{Cruise control reward ($r_2$)}
     \label{fig:cruise reward}
\end{subfigure}

\begin{subfigure}{\columnwidth}
\begin{tikzpicture}
\begin{axis}[
xtick={100000, 200000, 300000, 400000, 500000},
height=6cm,
width=8cm,
xmax=360000,
label style={font=\small},
ylabel={$r_{1,end}$},
xlabel={steps},
legend style={
legend cell align={left},
legend pos=south east,
font=\scriptsize
}
]
\addplot table [mark=none, x=Step, y=Smoothed Value, col sep=comma]{\rewardonedata};
\addplot table [mark=none, x=Step, y=Smoothed Value, col sep=comma]{\rewardonedatanine};
\addplot table [mark=none, x=Step, y=Smoothed Value, col sep=comma]{\rewardonedataninefive};
\addplot table [mark=none, x=Step, y=Smoothed Value, col sep=comma]{\rewardonedatamd};


\end{axis}
\end{tikzpicture}
     \centering
     \caption{Trj. control reward (at the end of episode)}
     \label{fig:trajectory reward}
\end{subfigure}
\vspace{-3mm}
\caption{The reward (the moving average with window size 10) achieved through fixed discount Q-learning and multi-discount Q-learning in I-AIM.}
\label{exp-mulit-discount}
\vspace{-4mm}
\end{figure*}

\clearpages

\bibliographystyle{named}
\bibliography{ijcai21}